\documentclass[12pt,preprint]{aastex}
\usepackage{tabularx}

\def\half{{\textstyle{1\over2}}}
 
\def\msun{{M_\odot}}

\def\Mt{M_{\rm t}}
\def\Cproj{C_{\rm proj}}

\def\vlos{{v_{\rm los}}}

\def\ds{\displaystyle}
\def\Rin{R_{\rm in}}
\def\Rout{R_{\rm out}}

\def\br{{\bf r}}
\def\bv{{\bf v}}

\def\rin{r_{\rm in}}
\def\rout{r_{\rm out}}
\def\sigmalos{\sigma_{\rm los}}
\def\kms{km\,s$^{-1}$}

\def\kpc{\ {\rm kpc}}

\def\spose#1{\hbox to 0pt{#1\hss}}
\def\lta{\mathrel{\spose{\lower 3pt\hbox{$\sim$}} \raise
2.0pt\hbox{$<$}}}
\def\gta{\mathrel{\spose{\lower 3pt\hbox{$\sim$}} \raise
2.0pt\hbox{$>$}}}

\begin{document}
\title{New Mass Estimators For Tracer Populations}
\author{N.W. Evans, M.I. Wilkinson}
\affil{Institute of Astronomy, Madingley Rd, Cambridge, CB3 0HA, UK}
\author{K.M. Perrett}
\affil{Department of Physics, Queen's University, Kingston,
Ontario, K7L 3N6, Canada}
\author{T.J. Bridges}
\affil{Anglo-Australian Observatory, Epping, NSW, 1710 Australia}
%
\begin{abstract}
We introduce the {\it tracer mass estimator}. This is a new and simple
way to estimate the enclosed mass from the projected positions and
line of sight velocities of a tracer population (such as globular
clusters, halo stars and planetary nebulae). Like the projected mass
estimator, it works by averaging (projected distance)$\times$ (radial
velocity)$^2/G$ over the sample. However, it applies to the commonest
case of all, when the tracer population does not follow the overall
dark matter density. The method is verified against simulated datasets
drawn from Monte Carlo realisations of exact solutions of the
collisionless Boltzmann equation and applied to the recent M31
globular cluster data set of Perrett et al. (2002), as well as to
M31's satellite galaxies.
\end{abstract}


\section{Introduction}

Simple estimators based on the virial theorem have long been used to
ascertain the masses of galaxy groups from radial velocities and
projected positions of the members (e.g. Zwicky 1933). Nowadays, they
have been largely superceded by the projected mass estimator, which is
based on averaging (projected distance)$\times$ (radial
velocity)$^2/G$ over the sample (Bahcall \& Tremaine 1981; Heisler,
Tremaine \& Bahcall 1985). In the form in which it is usually
encountered, a key assumption underlying the estimator is that the
member galaxies track the mass of the group.  The problem of
estimating the masses of elliptical galaxies and the haloes of spiral
galaxies is different. Here, the datasets available are the radial
velocities and projected positions of tracer populations, such as
dwarf galaxy satellites, globular clusters, distant halo stars and
planetary nebulae or PNe (see e.g., Wilkinson et al. 2002 for a recent
review). These do not follow the underlying mass distribution which is
dominated by the dark matter.  For nearby galaxies, the advent of
wide-field imaging and efficient planetary nebulae spectrographs
(e.g., M\'endez et al. 2001; Halliday et al. 2002) means that the
quality and quantity of such datasets are destined to improve
dramatically in the next few years.

The purpose of this paper is to present a new and simple mass
estimator for galaxy haloes tailored for tracer populations and to
apply them to new datasets for the Andromeda (M31) galaxy. It is the
generalisation of the projected mass estimator to the commonest case
of all -- namely that when the number density of the tracer population
is different from the overall mass density.

\section{Simple Mass Estimators}

On dimensional grounds, any mass estimator derived from projected data
on a set of $N$ tracer objects involves weighted means of $\vlos_i^2
R_i/ G$, where $\vlos_i$ is the line of sight velocity (relative to
the mean or systemic velocity) and $R_i$ the projected position of the
$i$th object relative to the center of the galaxy. The projected mass
estimator is
\begin{equation}
M = {\Cproj \over G N} \sum_i \vlos^2_i R_i.
\label{eq:proj}
\end{equation}
This was introduced by Bahcall \& Tremaine (1981) for test particles
orbiting point masses (such as stars around a black hole or distant
companion galaxies orbiting a central massive galaxy). The constant
$\Cproj = 16/\pi$ for test particles with an isotropic velocity
distribution orbiting a point mass, whereas $\Cproj = 32/\pi$ for test
particles moving on radial orbits.  Subsequently, Heisler et
al. (1985) extended the projected mass estimator to the case in which
the galaxy members are believed to track the total mass, such as
galaxy groups. The constant $\Cproj = 32/\pi$ for particles with an
isotropic velocity distribution, whereas $\Cproj = 64/\pi$ for
particles moving on radial orbits.

We wish to generalise the mass estimator (\ref{eq:proj}) to the case
when the objects are drawn from a tracer population (such as globular
clusters or PNe) between radii $\rin$ and $\rout$.  Let us assume that
the tracer population is spherically symmetric and has a number
density which falls off like a power-law
\begin{equation}
\rho (r) = \rho_0 \left( {a \over r} \right)^{\gamma},
\qquad\qquad \rin \lta r \lta \rout,
\label{eq:cuspdens}
\end{equation}
within the region of interest.  The underlying gravity field is
assumed to be scale-free
\begin{eqnarray}
\psi (r) =
\left\{ \begin{array}{ll}
{\ds v_0^2 \over \ds \alpha} \left({\ds  a \over \ds r} \right)^{\alpha} 
        &  \mbox{if $\alpha \neq 0$}, \\
\null & \null \\
-v_0^2 \log r  & \mbox{if $\alpha = 0$}.\end{array} \right.
\end{eqnarray}
When $\alpha =1$, this corresponds to test particles orbiting a
point-mass; when $\alpha =0$, the satellites are moving in a
large-scale mass distribution with a flat rotation curve; when $\alpha
= \gamma - 2$, the satellites follow the total self-gravitating mass
(the case studied by Heisler et al. 1985). Properties of such tracer
populations in scale-free spherical potentials have been deduced by
Evans, H\"afner \& de Zeeuw (1997).

Suppose our tracers are distributed over a range of projected radii
$\Rin$ and $\Rout$ (corresponding to the three-dimensional radii
$\rin$ and $\rout$). What mass distribution do they probe?  By
Newton's theorem, there is no knowledge whatsoever on the mass
distribution outside $\rout$.  Accordingly, the best we can do is to
estimate the total mass within the radius of the most distant satellite,
namely
\begin{equation}
M = {v_0^2 a^\alpha\over G} \rout^{1- \alpha},
\end{equation}
from the dataset of positions and velocities. Let ($r, \theta, \phi$)
be standard spherical polars with the $z$ axis defining the line of
sight, so that the projected position $R$ is $r\sin \theta$.  Denoting
the phase space distribution function of the tracer population by $f$,
then the average value of $\vlos^2 R$ in the survey is
\begin{equation}
\langle \vlos^2 R \rangle = {1\over \Mt} \int \int f \vlos^2 R d\bv
d\br = {4\pi \over \Mt} \int_{\rin}^{\rout} dr\int_0^{\pi/2} d
\theta\, \rho \sigmalos^2 r^3 \sin^2 \theta.
\label{eq:earlyone}
\end{equation}
Here, we have computed the average over the ensemble between the radii
$\rin$ and $\rout$ probed by our data. In this formula, $\sigmalos$ is
the line of sight velocity dispersion, while $\Mt$ is the mass in the
tracer population, which is given by
\begin{equation}
\Mt = \left\{ \begin{array}{ll}
{\ds 4 \pi a^\gamma \rho_0 \over \ds 3\!-\!\gamma} \left[ \rout^{3-\gamma}
- \rin^{3-\gamma}\right] & \mbox{if $\gamma \neq 3$}, \\
\null & \null \\
4 \pi a^3 \rho_0 \log (\rout/\rin) & \mbox{if $\gamma = 3$}.
\end{array} \right.
\label{eq:masst}
\end{equation}

\subsection{Isotropic Populations}

For the moment, let us assume that the velocity distribution of the
tracers is isotropic so that the radial velocity dispersion $\sigma_r$
is the same as the tangential $\sigma_t$. In this case, Evans et
al. (1997) solve the Jeans equations and derive the result
\begin{equation}
\rho \sigmalos^2 = \rho \sigma_r^2 = {\rho_0 v_0^2 \over \alpha
+\gamma} \left( {a \over r} \right)^{\alpha+\gamma}.
\label{eq:dispersions}
\end{equation}
For example, if the potential is isothermal ($\alpha =0$) with a flat
rotation curve of amplitude $v_0$ and the density of the population
falls off like $r^{-\gamma}$, then the velocity dispersion is just
$v_0/\sqrt{\gamma}$ -- which is a well-known result from the last
century (e.g., Smart 1938). Now, substituting (\ref{eq:masst}) and
(\ref{eq:dispersions}) into (\ref{eq:earlyone}), we obtain:
\begin{equation}
\langle \vlos^2 R \rangle = {\pi^2 \rho_0 v_0^2 \over \Mt (\alpha\!+\!\gamma)}
\int_{\rin}^{\rout} dr r^3 \left({a \over r} \right)^{\alpha+\gamma} =
{G M \pi \over 4 (\alpha\!+\!\gamma)}{ 3\!-\!\gamma \over
4\!-\!\alpha\!-\!\gamma}{ 1\!-\!(\rin/\rout)^{4-\alpha -\gamma} \over {
1\!-\!(\rin/\rout)^{3-\gamma}}}.
\label{eq:earlytwo}
\end{equation}
In other words, suppose we gather positions and velocities for a
population of globular clusters or PNe with a three-dimensional number
density falling like $r^{-\gamma}$ between an inner radius $\rin$ and
an outer radius $\rout$.  The mass enclosed within the outermost
datapoint is given by the simple formula
\begin{equation}
M = {C \over G N} \sum_i \vlos^2_i R_i,
\label{eq:massnew}
\end{equation}
with
\begin{equation}
C = {4(\alpha\!+\!\gamma) \over \pi}
    {4\!-\!\alpha\!-\!\gamma\over 3\!-\!\gamma}
    {1\!-\!(\rin/\rout)^{3-\gamma} \over
     1\!-\!(\rin/\rout)^{4\!-\!\alpha\!-\!\gamma}}.
\label{eq:massnewconst}
\end{equation}
We refer to this as {\it the tracer mass estimator}. At first sight,
it seems that the constant $C$ in eq~(\ref{eq:massnewconst}) is singular
when the tracer population density falls off like $r^{-3}$ (that is,
$\gamma =3$) or when $\alpha + \gamma =4$ (for example, if the tracer
population density falls off like $r^{-4}$ in an isothermal
($\alpha=0$) potential).  However, a careful application of
L'H\^opital's rule shows that the constants are well-defined. They are
listed in Table~\ref{tab:constants} for convenience.

How do we set the parameters occurring in eq~(\ref{eq:massnewconst})?
It is always reasonable to assume that $\rin \approx \Rin$, where
$\Rin$ is the projected radius of the innermost datapoint.  If the
dataset is derived from a wide-angle survey, in which the population
is traced out to large radii, then it is also reasonable to assume
that $\rout \approx \Rout$, the projected radius of the outermost
datapoint.  But, this assumption is not appropriate for a dataset
restricted to the inner parts only, because this will usually contain
objects at larger three-dimensional radii projected into the sample.
The parameter $\gamma$ can be calculated from the surface density of
the tracer population between $\Rin$ and $\Rout$.  However, $\alpha$
(and also possibly $\rout$) need to be set using astrophysical
considerations. We are envisaging applications to the outer parts of
galaxies probed by globular clusters and PNe, and so it is reasonable
to set $\alpha \approx 0$ as the potential is probably close to
isothermal. For example, Wilkinson \& Evans (1999) studied the Milky
Way galaxy and found that a near-isothermal potential is valid out to
$\sim 170$ kpc.

Comparison of the result (\ref{eq:massnew}-\ref{eq:massnewconst})
with previous work is instructive. For example, when $\alpha =1$, the
potential is Keplerian and we find that
\begin{equation}
M = {4(1\!+\!\gamma) \over \pi G N} \sum_i \vlos^2_i R_i.
\end{equation}
When $\alpha = \gamma -2$, the tracer population follows the
underlying dark matter density and we find that
\begin{equation}
M = {16(\gamma\!-\!1) \over \pi G N} {1 \over
\left[1\!+\!(\rin/\rout)^{3-\gamma}\right]} \sum_i \vlos^2_i R_i.
\end{equation}
These are not quite the same as the results given by Bahcall \&
Tremaine (1981) and Heisler et al. (1985).  These investigators
assumed that the sample was gathered from the center of the galaxy or
cluster to infinity. In a number of integration by parts, boundary
terms could therefore be legitimately dropped under the assumption
$r^3 \rho \sigma_r^2 \rightarrow 0$ as $r \rightarrow 0$ and as $ r
\rightarrow \infty$.  In our calculation, we have always performed the
averaging over a finite range of radii and the contribution from the
boundaries does not generally vanish. This is the origin of the
difference. Even when we formally take the limit $\rin \rightarrow 0$
and $\rout \rightarrow \infty$, the boundary terms still do not vanish
as the density distribution (\ref{eq:cuspdens}) of the tracer
population is cusped. Of course, such a limit is purely formal, as our
estimator is derived under the assumption that the power-law density
distribution holds over a certain r\'egime only.

\begin{table}[t]
\caption{The constant $C$ in the tracer mass estimator
eq~(\ref{eq:massnew}) for the two special cases ($\gamma =3$ and
$\alpha + \gamma =4$).  The number densities of tracer populations
belonging to the spheroid or stellar halo are often $\propto r^{-3}$
or $\propto r^{-4}$ in the outer reaches, so the special cases often
occur in practice.}
\label{tab:constants}
\begin{center}
\begin{tabular}{cc}
\hline
Case & Constant \\ \hline
\vspace{-.3cm}\\
$\gamma = 3$ & $C = {\ds 4(\alpha\!+\!3)(1\!-\!\alpha) \over \ds \pi} 
{\ds \log (\rout/\rin)\over \ds 1\!-\!(\rin/\rout)^{1\!-\!\alpha}}$ \\
\vspace{-.3cm}\\
$\alpha + \gamma =4$ & 
$ C = {\ds 16 \over \ds \pi  (3\!-\!\gamma)}
{\ds 1- (\rin/\rout)^{3-\gamma} \over \ds \log (\rout/\rin)}$ \\
\vspace{-.3cm}\\
\hline
\end{tabular}
\end{center}
\end{table}

\subsection{Anisotropic Populations}

Now let us extend the calculation to deal with populations with
anisotropic velocity distributions, although still retaining the
assumption of spherical symmetry for the density
distribution. Binney's (1981) anisotropy parameter $\beta = 1 -
\sigma_t^2/\sigma^2_r$ is often used to measure the relative
importance of the radial $\sigma_r$ and tangential $\sigma_t$ velocity
dispersions. When $\beta \rightarrow - \infty$, this is the circular
orbit model and when $\beta =1$, this is the radial orbit model.  For
models in which the anisotropy $\beta$ does not vary with radius, the
line of sight velocity dispersion $\sigmalos$ is related to the radial
velocity dispersion by
\begin{equation}
\sigmalos^2 = \sigma_r^2 ( 1- \beta \sin^2 \theta).
\end{equation}
Substituting into (\ref{eq:earlyone}), we find
\begin{equation}
\langle \vlos^2 R \rangle
= {\pi^2 (4\!-\!3\beta) \over 4\Mt} \int_{\rin}^{\rout} 
\rho \sigma_r^2 r^3 dr.
\end{equation}
The radial velocity dispersion $\sigma_r$ of the tracer population is
(Evans et al. 1997)
\begin{equation}
\rho \sigma_r^2 = {\rho_0 v_0^2 \over \alpha\!+\!\gamma\!-\!2\beta} \left(
{a \over r} \right)^{\alpha+\gamma}.
\end{equation}
From this, we deduce
\begin{equation}
M = {C \over G N} \sum_i \vlos^2_i R_i,
\label{eq:anisoa}
\end{equation}
with
\begin{equation}
C = {16(\alpha\!+\!\gamma\!-\!2\beta) \over \pi (4\!-\!3\beta)}
    {4\!-\!\alpha\!-\!\gamma \over 3\!-\!\gamma}
    {1\!-\!(\rin/\rout)^{3-\gamma} \over
     1\!-\!(\rin/\rout)^{4-\alpha -\gamma}}
\label{eq:anisob}
\end{equation}
This is the extension of the result to populations with constant
anisotropy. Unless there are compelling reasons to the contrary (e.g.,
a highly flattened system supported by an anisotropic velocity
dispersion), we advocate assuming isotropy. The value of
eq~(\ref{eq:anisoa}) is that it enables us to calculate the error in
making such an approximation. For example, suppose we consider tracer
populations with a number density falling like $r^{-4}$ in an
isothermal potential. The anisotropy of stellar populations is not
usually more extreme than $2:1$. Therefore, eq~(\ref{eq:anisoa}) tells
us that mistakenly assuming isotropy leads to mass underestimates
(overestimates) in the case of radial (tangential) anisotropy of $
\sim 30 \%$.

\begin{table}[t]
\caption{The percentage of mass estimates in error by a large factor
(too big or too small by 50 per cent). The numbers are derived from
10000 Monte Carlo simulations of isotropic tracer populations with a
number density falling like $r^{-4}$ in an isothermal potential. The
sample size $N$ and extent $\rout/\rin$ are chosen so as to be typical
of datasets of satellite galaxies (upper panel) and globular clusters
(lower panel).}
\label{tab:percents}
\begin{center}
\begin{tabular}{ccccc}
\hline
Estimator & Number & $\rout/\rin$ & Too Low & Too High  \\ \hline
Tracer & 10 & 100 & $20 \%$ & $ 37\%$ \\
Projected & 10 & 100 & $ 53\%$ & $11 \%$ \\
Virial & 10 & 100 & $87 \%$ & $ 0\%$ \\
\hline
Tracer & 100 & 10 & $ 2\%$ & $2 \%$ \\
Projected & 100 & 10 & $72 \%$ & $0 \%$ \\
Virial & 100 & 10 & $100 \%$ & $ 0 \%$ \\
\hline
\end{tabular}
\end{center}
\end{table}

\subsection{Monte Carlo Simulations}

We test the performance of the mass estimator with Monte Carlo
realisations of exact solutions of the collisionless Boltzmann
equation. The phase space distribution functions corresponding to
power-law density profiles in power-law potentials are given in Evans
et al. (1997). To generate Monte Carlo realisations, the speed is
picked from the distributions
\begin{eqnarray}
f (v) \propto
\left\{ \begin{array}{ll}
v^{2 - 2\beta} \left|\psi(r) - \half v^2 \right|^{[2\gamma - 3\alpha
- 2\beta(2-\alpha)]/(2\alpha)}
&  \mbox{if $\alpha \neq 0$}, \\
\null & \null \\
v^{2- 2\beta} \exp \left( -{\ds v^2 \over \ds 2\sigma^2} \right)
& \mbox{if $\alpha = 0$}.\end{array} \right.
\end{eqnarray}
For $\alpha >0$, the maximum velocity at any position is $\sqrt{2
\psi(r)}$; for $\alpha \le 0$, the velocities can become arbitrarily
large.  Following Binney \& Tremaine (1987), let us introduce
spherical polar coordinates in velocity space ($v, \xi, \eta)$, so
that the velocities resolved in spherical polar coordinates with
respect to the center are then
\begin{equation}
v_r = v \cos \eta,\quad v_\theta = v \sin \eta \cos \xi, \quad
v_\phi = v \sin \eta \sin \xi.
\end{equation}
If $\xi$ is generated uniformly in $[0,2\pi]$, while $\eta$ is picked
in $[0,\pi]$ from the distribution
\begin{equation}
F(\eta) \propto | \sin \eta |^{1 + 2\beta}
\end{equation}
then the velocities are generated with the correct anisotropy.
Finally, a random projection direction is chosen and the line of sight
velocity $\vlos$ and projected position $R$ calculated. This gives us
synthetic datasets of anisotropic spherical tracer populations in
spherical haloes with which to test the performance of our mass
estimators.

\begin{figure}
\begin{center}
\plotone{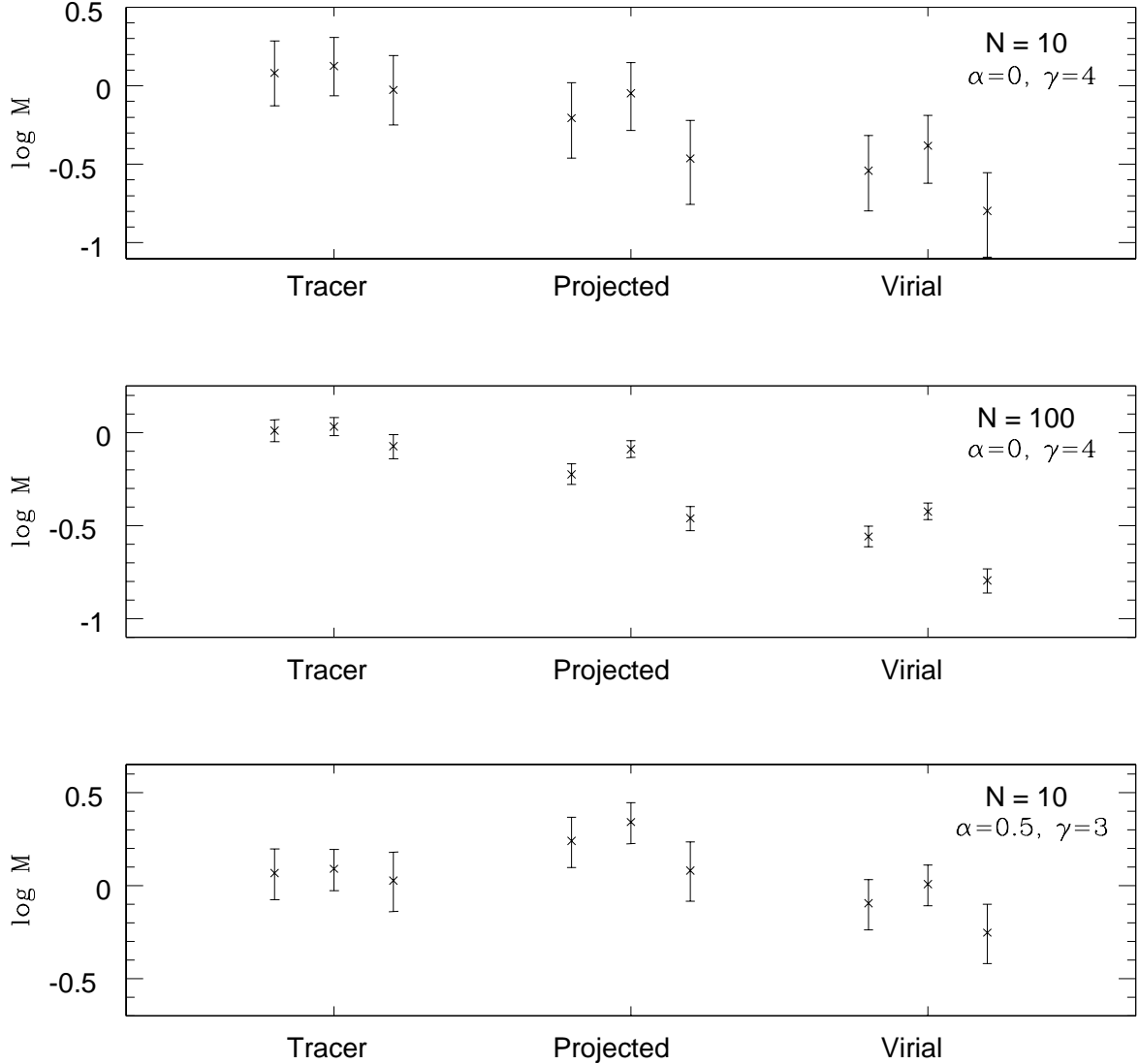}
\end{center}
\caption{Medians and upper and lower quartiles for the mass estimates
of 10000 tracer samples constructed from realisations of exact
scale-free models.  Upper Panel: The tracer population has a number
density falling like $r^{-4}$ in an isothermal potential. The data is
gathered between $\Rout = 10\Rin$. From left to right, the models are:
(i) 10 objects drawn from an isotropic velocity distribution, (ii) 10
objects drawn with $\sigma_t:\sigma_r = 2:1 $, and (iii) 10 objects
with $\sigma_t:\sigma_r = 1:2$.  For anisotropic populations, knowledge
of the anisotropy $\beta$ is assumed and the tracer mass estimator in
the general form (\ref{eq:anisoa}) is used.  Middle Panel: As the
Upper Panel, but the number of objects is now 100. Lower Panel: As the
Upper Panel, but the tracer population has a number density falling
like $r^{-3}$ in a potential with a falling rotation curve $\psi
\propto r^{-1/2}$.}
\label{fig:montecarlos}
\end{figure}

Figure~\ref{fig:montecarlos} shows the medians and quartiles for the
mass estimates of 10000 samples constructed from Monte Carlo
realisations. We test the tracer mass estimator for both isotropic
(eq.~\ref{eq:massnew}) and anisotropic (eq.~\ref{eq:anisoa})
populations, as well as the projected and virial mass estimators in
the forms
\begin{equation}
M_{\rm proj} = {32 \over \pi G N} \sum_i \vlos^2_i R_i,\qquad\qquad
M_{\rm virial} = {3\pi N \over 2G} { \sum_i \vlos^2_i \over \sum_{i} 1/ R_i}.
\label{eq:virial}
\end{equation}
For each of the mass estimators, four models are shown in each
panel. In the upper two panels, the tracer population has a number
density falling like $r^{-4}$ and is generated between $\Rin$ and
$\Rout = 10 \Rin$, while the rotation curve is flat ($\alpha =0$), The
models vary in the anisotropy of the velocity distribution (either
$\sigma_t: \sigma_r = 1:1$ or $\sigma_t:\sigma_r = 2:1 $ or
$\sigma_t:\sigma_r = 1:2$). In upper panel, the number of objects is
$N=10$, which is typical for a sample of satellite galaxies.  In the
middle panel, the number of objects is $N=100$, which is realistic for
a sample of globular clusters with radial velocities.  Finally, in the
lower panel, the rotation curve is falling in a ``half-way Keplerian''
manner ($\alpha = 1/2$) and the tracer population is falling like
$r^{-3}$. The sample size is again $N =10$.  The tracer mass estimator
outperforms the projected and virial mass estimators in every
case. This is as it should be, since the tracer mass estimator has
been devised for the precise purpose of mass estimation from tracer
populations, while the other two estimators have been devised with
other applications in mind. It is interesting to ask how often the
estimates are too low or too high by $50 \%$. This information is
recorded in Table~\ref{tab:percents} for some of the simulations. The
projected and virial mass estimators yield systematic underestimates
if they are mistakenly applied to tracer populations. Further, the
situation is not improved by increasing the size of the tracer
population -- the estimators merely converge to an underestimated
mass.

\begin{figure}
\begin{center}
\plotone{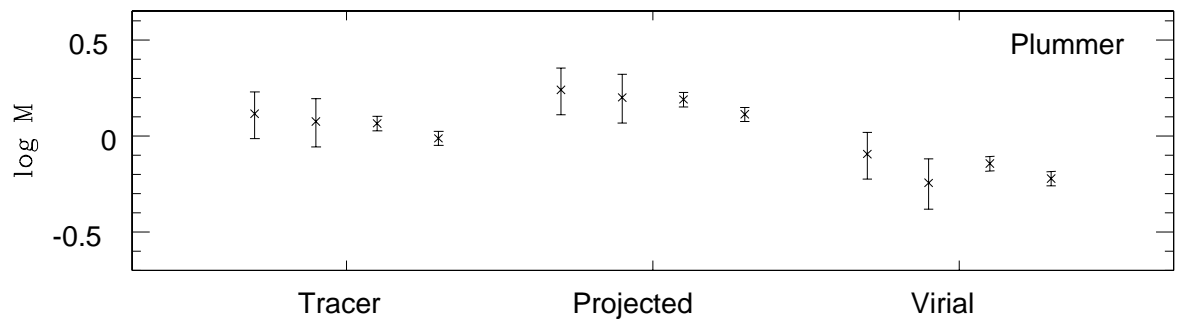}
\end{center}
\caption{Medians and upper and lower quartiles for the mass estimates
of 10000 tracer samples constructed from realisations of exact Plummer
models with core radius $b$. The self-consistent population has a
number density falling like $r^{-5}$ in a potential falling like
$r^{-1}$ at large radii.  Again, the performance of the tracer,
projected and virial mass estimators is compared.  From left to right,
the models are: (i) 10 objects gathered between $\Rin = 10 b$ and
$\Rout = 100b$, (ii) 10 objects between $\Rin = b$ and $\Rout = 100b$,
(iii) the same as (i) but with $N=100$, and (iv) the same as (ii) but
with $N=100$.}
\label{fig:plums}
\end{figure}

Finally, we test the tracer mass estimator against synthetic datasets
drawn from a self-consistent Plummer model with a core radius
$b$. This is a important thing to do as the tracer estimator has been
derived for, and tested against, the scale-free case. The Plummer
model is only approximately scale-free well outside the core region ($
r \gg b$). The potential, density and distribution of speeds are
\begin{equation}
\rho \propto {1 \over (r^2 + b^2)^{5/2}}, \qquad \psi \propto {1 \over 
(r^2 + b^2)^{1/2}}, \qquad f(v) \propto v^2 (\psi(r) - \half v^2)^{7/2}. 
\end{equation}
Figure~\ref{fig:plums} shows how the estimators fare for sample sizes
$N=10$ and $N=100$ using datasets gathered in the outer parts ($\Rin =
10b, \Rout = 100b$) and including some core contamination ($\Rin = b,
\Rout = 100b$).  For a self-consistent population, the projected mass
estimator performs well, as it should do. Nonetheless, the tracer mass
estimator (with $\gamma$ = 5, $\alpha$ =1 and $\beta$ = 0) gives
slightly better results. The reason why the tracer mass estimator
still out-performs the projected mass estimator is that the former
includes a correction for the limited radial coverage of the sample,
whereas the latter does not.

\section{Application: the Andromeda Galaxy}

It is important to know the mass of the Andromeda Galaxy (M31) for a
number of reasons. First, the density profiles of dark matter haloes
are an important constraint on cosmological theories of galaxy
formation (e.g., Sellwood 2001). The haloes of nearby galaxies are
particularly important as they can be studied in much greater detail.
Second, the current generation of pixel lensing experiments towards
M31 (e.g., Auri\`ere et al. 2001, Paulin-Henriksson et al. 2002)
requires an accurate estimate of the total mass so as to constrain the
fraction in compact objects capable of producing microlensing events.
We shall consider two datasets which probe Andromeda's dark halo,
namely the globular clusters and the dwarf spheroidal satellite
galaxies.

\subsection{Globular Clusters}

\begin{figure}
\begin{center}
\plotone{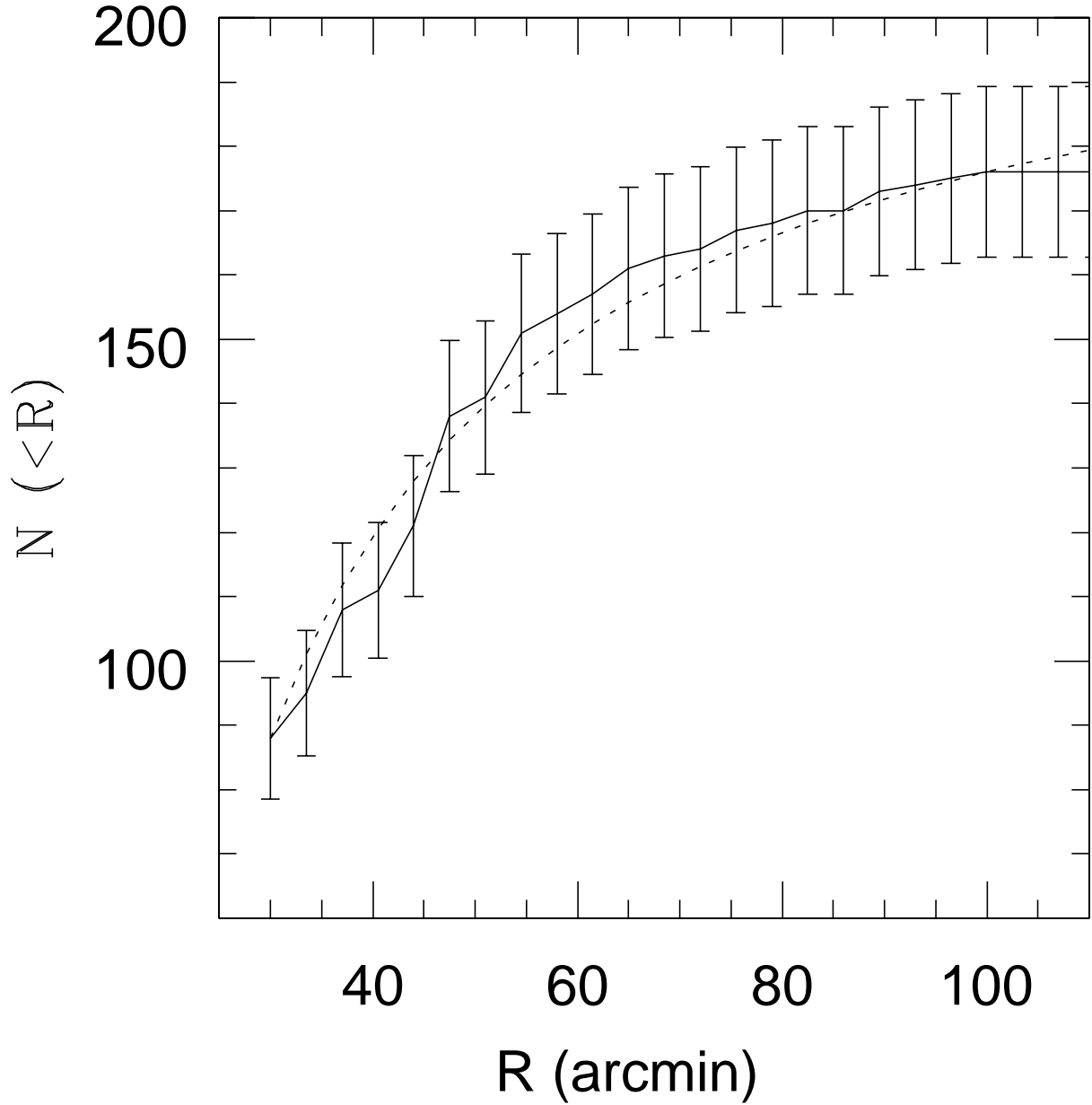}
\end{center}
\caption{Cumulative number $N$ of M31 halo globular clusters versus
projected radius $R$ (in arcmin). The dotted line shows the profile
expected for a population with three dimensional density distribution
$\rho \sim r^{-4}$ between $30^\prime$ and $100^\prime$.}
\label{fig:GC_N_vs_R}
\end{figure}

Perrett et al. (2002) give the positions, velocities and metallicities
for over 200 globular clusters in M31 out to a projected distance
$\sim 30$ kpc. The data were obtained using fiber optic spectroscopy
on the {\it William Herschel Telescope} and the typical errors on the
radial velocities are $\pm 12$ \kms.  When combined with earlier data
from Huchra, Brodie \& Kent (1991), this gives a grand total of $\sim
300$ globular clusters.  From the bimodality of the metallicity
distribution, Perrett et al. show that the M31 globular clusters fall
into metal-poor and metal-rich categories with different kinematics.
As judged from the double Gaussian fit to the metallicity
distribution, there are 177 globular clusters with a probability of
belonging to the halo of greater than $90 \%$.  The cumulative number
distribution of the halo globular clusters is shown in
Figure~\ref{fig:GC_N_vs_R}.  Only beyond 30 arcmin does the projected
number density of the halo globulars fall off like a power-law, namely
$R^{-3}$.  Accordingly, we work with the 89 halo globular clusters
with projected radii greater than 30 arcmin. The mean rotation
amplitude of this sub-sample is $\langle v_\phi \rangle \approx 110$
\kms. Adopting a distance of 770 kpc, the globular clusters lie in
projection between 6.8 kpc and 33 kpc. However, the globular cluster
population in M31 certainly extends out to $\sim 100$ kpc (e.g., Hodge
1992). Using Monte Carlo simulations to build samples of 88 globular
clusters with $6.8 \kpc <R< 33 \kpc$, we find that typically $20 \%$
of the sample have $r > 33 \kpc$ with the outermost datapoints
typically at $r \approx 100$ kpc.  Accordingly, in the formula for the
tracer mass estimator, we set $\alpha = 0$ (isothermal-like galaxy),
$\gamma \approx 4$, $\rin = 6.8$ kpc and $\rout \approx 100$ kpc.

The Jeans equation for a population with constant anisotropy $\beta$
about a mean velocity $\langle v_\phi \rangle$ is (e.g., Binney \&
Tremaine 1987, section 4.2)
\begin{equation}
{GM(r) \over r} = \langle v_\phi \rangle^2 - \sigma_r^2 \left[
{\partial \log \rho \sigma_r^2 \over \partial \log r} + 2 \beta \right]
\end{equation}
The first term on the right-hand side describes the contribution of
rotation to $M(r)$, the second term the contribution of pressure.
With obvious notation, we write
\begin{equation}
{GM_{\rm rot} \over r} = \langle v_\phi \rangle^2, \qquad\qquad
{GM_{\rm press} \over r} = - \sigma_r^2 \left[ {\partial \log \rho \sigma_r^2
\over \partial \log r} + 2 \beta \right]
\end{equation}
We compute $M_{\rm rot} \sim 3 \times 10^{11} \msun$ directly from the
mean rotation amplitude of the sample. We use the tracer mass
estimator, applied to the observational data with the mean rotation
velocity subtracted from the line of sight velocity, to compute
$M_{\rm press} \sim 9 \times 10^{11} \msun$ assuming isotropy. Hence,
the total mass within $\sim 100$ kpc of the center of M31 is $\sim 1.2
\times 10^{12} \msun$.  Note the tracer mass estimator applies to a
pressure-supported tracer population and therefore any net rotation of
the system must be subtracted before the velocities are used to
estimate $M_{\rm press}$.

How does the result from the tracer mass estimator compare with
earlier work?  Evans \& Wilkinson (2000) give the results of more
sophisticated modelling. Here, the phase space distribution function
of the globular cluster and the satellite galaxy population is built
and then convolved with the errors to find the probability of the
dataset given the model parameters.  This is inverted using Bayesian
likelihood techniques to give the probability of the enclosed mass
given the positions and velocities of the halo globular
clusters. Using the combined globular cluster and satellite galaxy
sample, they reckoned that the most likely total mass of M31 is $\sim
12^{+18}_{-6} \times 10^{11} \msun$.  An independent constraint on the
mass of M31 within 30 kpc is given by the HI rotation curve. This was
measured to distances of $\sim 30$ kpc along the major axis by Newton
\& Emerson (1977). Their estimate of the asymptotic circular speed is
230 \kms.  Assuming the circular velocity curve maintains this
amplitude out to 100 kpc, then the mass within 100 kpc is $1.2 \times
10^{12} \msun$. We conclude that the results are all consistent and
support a picture in which M31's dark halo extends out in an
isothermal-like manner to at least 100 kpc.

\subsection{Satellite Galaxies}

There are 15 probable companion galaxies to M31. Their galactic
coordinates, projected position from the center of M31 and line of
velocity are listed in Table~\ref{tab:satdata}.  The observed radial
velocities contain contributions from the Galactic rotation and from
the relative velocity between the Galaxy and M31. Assuming that the
transverse velocity of M31 with respect to the Galaxy is zero, we
remove these contributions and list the corrected velocities in the
frame of M31 in Table~\ref{tab:satdata}.

The number density of M31's satellite galaxies falls like $r^{-3.5}$
(see e.g., Evans \& Wilkinson 2000).  Notice that the projected
positions of many of the objects lies well beyond the r\'egime in
which the M31 halo is expected to be isothermal. Rather, the
underlying gravitational potential is probably dominated by the
monopole component at such huge distances. This suggests that it is
appropriate to apply the tracer mass estimate with $\alpha \approx 1$
and $\gamma \approx 3.5$ to the satellite galaxy dataset.  This gives
the mass of M31 as $\sim 1.1 \times 10^{12} \msun$.  This is in good
agreement with the more sophisticated modelling in Evans et
al. (2000), who claimed that the total mass is $7-10 \times 10^{11}
\msun$ from the same dataset. In fact, it is in excellent accord with
still earlier work by Bahcall \& Tremaine (1981), who also found a
mass of $\sim 1.0 \times 10^{12} \msun$ from just five companions
(M32, M33, NGC 147, NGC 185 and NGC 205).  Despite the seeming
agreement, the uncertainty in the mass is at least a factor of 3.  For
example, if instead we choose $\alpha \approx 0.5$ (suitable for a
halo with a rotation curve slowly falling like $\propto r^{-0.25}$),
then the mass of M31 as deduced from the tracer estimator more than
triples to $\sim 3.8 \times 10^{12} \msun$.

An assumption underlying the tracer mass estimator (as well as the
projected and virial estimators) is that of a steady state
equilibrium. This is reasonable enough for globular cluster and PNe
datasets, but probably not for satellite galaxies.  The dSph and dIrr
companions of M31 may well be falling in for the first time, in which
case the assumption of a steady state is probably not a valid
description of the dynamics.  It would be valuable to calibrate the
importance of the effects of disequilibrium against high resolution
N-body simulations of the Local Group.

\begin{table}[t]
\caption{Data on the companion galaxies of M31 taken from Grebel
(2000) and Evans et al. (2000).  Listed are Galactic coordinates
($\ell$,$b$), the projected distance from the center of M31 $R$ in
kpc, corrected line of sight velocities $\vlos$ (adjusted for the
solar motion within the Galaxy and the radial motion towards M31)
and object type. (Note: the coordinates for the Cas\,dSph given in
Table~1 of Evans et al. (2000) are incorrect).}
\label{tab:satdata}
\begin{center}
\begin{tabular}{rrrrrc}
\hline\hline
Name& \multicolumn{1}{c}{$\;\ell$}& \multicolumn{1}{c}{$\;\;b$}&
\multicolumn{1}{c}{$\;R$} 
& \multicolumn{1}{c}{$\vlos$} & Type\\ \hline
\vspace{-.3cm}\\
M31 & 121.2 & $-21.6$& $-$& $-$& SbI-II\\
\vspace{-.3cm}\\
\hline
\vspace{-.3cm}\\
M32    & 121.1 & $-22.0$ & 5 & $+95$ & E2 \\
NGC 205 & 120.7 & $-21.1$ & 8 & $+58$ & dSph \\
NGC 147 & 119.8 & $-14.3$ & 100 & $+118$ & dSph/dE5 \\
NGC 185 & 120.8 & $-14.5$ & 95 & $+107$ & dSph/dE3 \\
M33 & 133.6 & $-31.5$ & 198 & $+72$ & ScII-III \\
IC 10 & 119.0 & $-3.3$ & 243 & $-29$ & dIrr \\
LGS3 & 126.8 & $-40.9$ & 262 & $-38$ & dIrr/dSph \\
Pegasus & 94.8 & $-43.5$ & 397 & $+86$ & dIrr/dSph \\
IC 1613 & 129.7 & $-60.6$ & 489 & $-58$ & IrrV \\
And\,I & 121.7 & $-24.9$ & 44 & $-85$ & dSph\\ 
And\,II & 128.9 & $-29.2$ & 138 & $+82$ & dSph\\ 
And\,III & 119.3 & $-26.2$ & 67 & $-58$ & dSph\\ 
And\,V & 126.2 & $-15.1$ &07 & $-107$ & dSph\\ 
And\,VI & 106.0 & $-36.3$ & 260 & $-65$ & dSph\\ 
Cas\,dSph & $109.5$ & $-9.9$ & 215 & $+23$ & dSph\\ 
\vspace{-.3cm}\\
\hline\hline
\end{tabular}
\end{center}
\end{table}

\section{Conclusions}

The main accomplishment of this paper is the introduction of the {\it
tracer mass estimator}. This is a new and simple way of estimating the
enclosed mass from the radial velocities and projected positions of a
set of objects. It is tailored for the case of tracer populations such
as globular clusters and halo stars whose number density falls off
like a power-law with distance (at least within the range covered by
the survey).  The estimator works by averaging the quantity (constant)
$\times$ (projected distance)$\times$ (radial velocity)$^2/G$ over the
sample. The value of the constant is straightforward to calculate. It
depends on the radial fall-off of the tracers, the inner and outer
radii marking the range of the sample, the velocity anisotropy of the
population and the underlying gravitational potential. In the absence
of further information, we recommend assuming isotropy and
isothermality for estimating galaxy masses, at least in the case of
globular cluster and planetary nebulae (PNe) datasets. The inner and
outer radii can usually be inferred from the sample itself, sometimes
using additional astrophysical evidence.

We have applied the tracer mass estimator to the globular cluster and
satellite galaxy datasets of the Andromeda galaxy (M31). The estimates
suggest a picture in which M31's dark halo is isothermal out to at
least $\sim 100$ kpc and has a mass of $\sim 1.2 \times 10^{12}
\msun$. These estimates are in good agreement with the results of more
sophisticated modelling. The advantage of the tracer mass estimator is
that it is so much quicker to apply and requires much less effort.  It
should be used in preference to the projected mass estimator for
populations in the outer parts of galaxies. One of the fundamental
assumptions of the projected mass estimator (at least in the form
which it is usually encountered) is that the population shadows the
overall mass density. This assumption breaks down for populations in
the outer parts of galaxies. We envisage the tracer mass estimator as
being particularly useful for early-type galaxies for which there is
no gas rotation curve and so mass estimates almost always derive from
dynamical modelling of tracer populations.

\acknowledgements NWE thanks the Royal Society and MIW thanks PPARC
for financial support. We are grateful to the (anonymous) referee for
providing a helpful report.

\end{document}